\documentclass[reprint,pra,superscriptaddress,nofootinbib,aps]{revtex4-1}
\usepackage{amsmath, amssymb}	% Loads AMS packages
\usepackage{bbm}				% Allows for bold Greek characters
\usepackage{braket}				% Allows Dirac notation
\usepackage{graphicx} 			% Allows for insertion of figures
\usepackage{mathtools}			% Use for small matrices
\usepackage{suffix}				% Allows for defining * version of custom commands
\usepackage{pst-math, pst-plot}	% Allows for post script pictures
\usepackage{mathtools}
\usepackage{hyperref}			% Hyper reference
\usepackage{cleveref}
\usepackage{bm}
\usepackage[T1]{fontenc}
\usepackage{parskip}
\newcommand{\avg}[1]{\left< #1 \right>} 				% for average
\DeclareMathOperator\arctanh{arctanh} 				% Inverse hyperbolic tangent
\renewcommand{\Re}{\mathop{\rm Re}}				    % for the real part
\renewcommand{\Im}{\mathop{\rm Im}}					% for the imaginary part
\newcommand{\nn}{\nonumber}

\def\eu{\ensuremath{\mathrm{e}}}
\def\iu{\ensuremath{\mathrm{i}}}
\usepackage{soul}
\begin{document}

\preprint{APS/123-QED}

\title{Relativistic (2,3)-threshold quantum secret sharing}

\author{Mehdi Ahmadi}
\email[]{mehdi.ahmadi@ucalgary.ca}
\affiliation{Department of Mathematics and Statistics, University of Calgary, Calgary, Alberta T2N 1N4, Canada} 
\affiliation{Institute for Quantum Science and Technology, University of Calgary, Alberta T2N 1N4, Canada}

\author{Ya-Dong Wu}
\email[]{yadong.wu@ucalgary.ca}
\affiliation{Institute for Quantum Science and Technology, University of Calgary, Alberta T2N 1N4, Canada}

\author{Barry C. Sanders}
\email[]{sandersb@ucalgary.ca}
\affiliation{Institute for Quantum Science and Technology, University of Calgary, Alberta T2N 1N4, Canada}
\affiliation{Program in Quantum Information Science, Canadian Institute for Advanced Research,Toronto, Ontario M5G 1Z8, Canada}
\affiliation{Hefei National Laboratory for Physical Sciences at Microscale,University of Science and Technology of China, Hefei, Anhui 230026, China}
\affiliation{Shanghai Branch, CAS Center for Excellence and Synergetic Innovation Center in Quantum Information and Quantum Physics, University of Science and Technology of China, Shanghai 201315, China}

\date{\today}	% It is always \today, today,

\begin{abstract}
In quantum secret sharing protocols, the usual presumption is that the distribution of quantum shares and players' collaboration are both performed inertially. Here we develop a quantum secret sharing protocol that relaxes these assumptions wherein we consider the effects due to the accelerating motion of the shares. Specifically,  we solve the $(2, 3)$-threshold  continuous-variable quantum secret sharing in non-inertial frames.
To this aim, we formulate the effect of relativistic motion on the quantum field inside a cavity as a bosonic quantum Gaussian channel. We  investigate how the fidelity of quantum secret sharing is affected by non-uniform motion of the quantum shares. Furthermore, we fully characterize the canonical form of the Gaussian channel which can be utilized in quantum information processing protocols to include relativistic effects.\end{abstract}

\maketitle

%========================================
%========================================

\section{Introduction}
\label{Introduction}
Continuous-variable quantum secret sharing is experimentally feasible~\cite{Lance1}, however, a comprehensive theory of continuous-variable quantum error correction is still missing; Gaussian states cannot be protected against Gaussian errors using Gaussian operations~\cite{Niset}. In previous studies~\cite{Tyc1,Lance2,Lance1}, the effect of non-inertial motion during the transmission of quantum shares has been ignored. Here we solve continuous-variable quantum secret sharing wherein the quantum shares move non-uniformly in Minkowski spacetime and our results show how acceleration affects the fidelity of quantum secret sharing.\\

In $(k, n)$-threshold quantum secret sharing, the dealer encodes a quantum secret in $n$ quantum systems (or quantum shares), which he then distributes to $n$ players. Each player receives exactly one share, where any subsets of $k$ or more players form the access structure to retrieve the secure key while any subsets of fewer than $k$ players; i.e., the adversarial structure, cannot learn any information whatsoever about the key. Continuous-variable threshold quantum secret sharing still faces the challenge that information about the quantum secret can be leaked into the adversarial structure~\cite{Tyc2,Kogias}. Various models
of secret sharing exist with quantum or classical channels that can
be public or private and a graph-state formalism was proposed to unify these models~\cite{Markham2008}. Here we consider the scenario wherein the dealer shares quantum
channels with each player, and also the players share quantum channels between each other, which is known as the QQ case~\cite{Markham2008}.\\

We focus on a relativistic variant of a $(2,3)$-threshold quantum secret sharing protocol which is the smallest-sized non-trivial protocol. We take into account the relativistic motion of the quantum shares in Minkowski spacetime during the distribution and collaboration and how it influences the success of the protocol. In our relativistic protocol, similar to the non-relativistic case~\cite{Cleve}, a dealer encodes the quantum secret into several quantum shares and distributes them to all the players. The players are located at different regions in the Minkowski spacetime and the dealer and the players are all stationary. Under such circumstances, during the dealer's distribution, the quantum shares experience non-uniform motion, as they are transmitted to spacetime points in the future light cone of the dealer (Fig.~\ref{fig:distribution}). Then, a subset of players within the access structure collaborate to retrieve the quantum secret by sharing their individual shares. However, to reach the same spacetime point, the shares go through phases of accelerating and decelerating motion while being transmitted. We analyze the possible collaboration scenarios between the players; i.e., Players~1 and~2 collaborate (Fig.~\ref{fig:collaboration12}) or Players~2 and~3 collaborate (Fig.~\ref{fig:collaboration23}). In each scenario, we investigate how the non-inertial motion of the shares affects the fidelity of the quantum secret sharing protocol.\\

In $(2,3)$-threshold quantum secret sharing, the dealer encodes the quantum secret in three quantum shares in a localized manner, hence, we need to be able to analyze the effect of relativity on such systems. The relativistic effects on the state of localized quantum systems has been studied using different setups~\cite{Bruschi2012,Friis,Dragan1,Dragan2,Ahmadi2016,Richter2017,Downes2013}. We find the framework of accelerating cavities a suitable choice for this purpose, as it can be adapted to study the effect of non-uniform motion on localized quantum fields~\cite{Bruschi2012,Friis}. Accelerating cavities have been employed in the past to study the relativistic effects on quantum clocks~\cite{Lindkvist}, quantum teleportation~\cite{Nico2013}, and to estimate proper acceleration~\cite{Ahmadi2014a,Ahmadi2014b} to name a few. However, we develop a different approach from the previous studies for accelerating cavities; we formulate the evolution of the quantum field inside an accelerating cavity as a bosonic quantum Gaussian channel (BQGC) which we then use to include the effects of non-uniform motion of the quantum shares. Furthermore, this approach enables us to compute physical quantities, such as the average number of produced thermal particles and transmissivity of the relativistic channel.\\ 
 
This paper is organized as follows. In Section \ref{Background}, we provide a brief review of how a quantum field inside a cavity is affected due to relativistic motion. In Section \ref{Methods}, we formulate the change in the state of the quantum field as a BQGC and utilize the canonical form of the BGQC to study particle creation inside the cavity and transmissivity of the channel. In Section \ref{Results}, we employ the channel to study the effect of relativity on the $(2,3)$-threshold continuous-variable quantum secret sharing for different collaboration scenarios and also for different quantum Gaussian secret states. Finally, in Section \ref{Discussion}, we discuss our results and provide future lines of research. Throughout this paper, we use units in which $c=\hbar=1$.

%%%%%%%%%%%%%%%%%%%%%%%%%%%%%%%%%%%%%%%%%%%%%%%%%%%%%%%%%%%%%%%%%%%%%
\section{Background}\label{Background}
\begin{figure}
   \centering
   \includegraphics[width=0.9\linewidth]{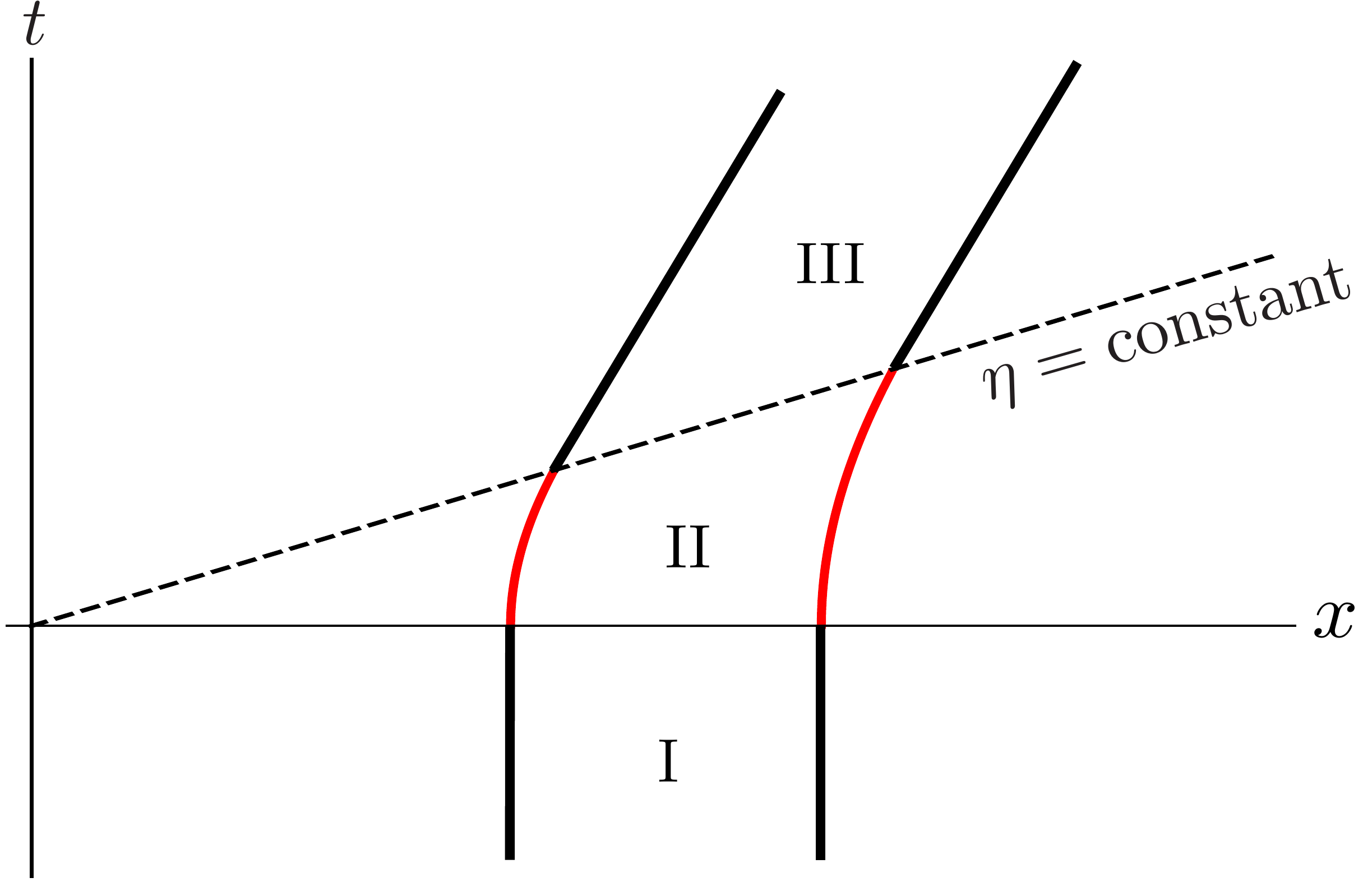} 
   \caption{Basic Building block for an arbitrary trajectory. The world lines of the left and right walls of the cavity are depicted. In region I, the cavity is inertial. In region II, the two walls of the cavity are accelerating with two different proper accelerations until the Rindler coordinate time $\eta=\frac{\tau}{a}$, where $\tau$ and $a$ are proper time and acceleration respectively. In region III, the cavities have stopped accelerating and back in the inertial frame again. The hyperbolas (red curves) represent the trajectories of the cavity walls moving with constant proper acceleration, and the (black) straight lines correspond to the trajectories of the walls while they move inertially.}
   \label{BBB}
\end{figure}
In this section, we briefly review the effect of acceleration on the evolution of a quantum field inside a cavity, which is a well-studied topic (we refer the interested reader for more details to~\cite{Friis}). We focus on a simple trajectory which is known in the literature as the \textsl{basic building block} (BBB), since it enables studying any arbitrary non-uniform trajectory~\cite{Friis}. As depicted in Fig.~\ref{BBB}, the BBB employs three steps. Initially, in region I, the cavity is at rest. Then, it accelerates for some time in region II and finally, in region III, it goes back to being inertial again. We note that, for the cavity to remain rigid\footnote{The cavity motion is constructed so that the cavity remains rigid. This is in the sense that a comoving observer sees the cavity walls at a constant proper distance at all times.}, different parts of it need to accelerate at different rates~\cite{Friis} (as shown in Fig.~\ref{BBB}).\\

The inertial to uniformly accelerated transition can be modelled as a unitary linear transformation of the mode operators~\cite{Friis}. As there is a unique correspondence between any such unitary and a symplectic transformation on the phase space~\cite{Gerardo2014} , we represent the transformation of the quantum field inside the cavity from region I to region II by the symplectic transformation $S_{\text{\tiny I,II}}$. A symplectic transformation is a transformation that preserves the symplectic form; i.e., 
\begin{align}\label{Sym}
	S\Gamma S^{T}=\Gamma\,;\,\,\,
\Gamma :=\bigoplus_{i} \Gamma_i, \,\,\, \Gamma_i :=\begin{bmatrix} 0 & 1\\ -1 & 0 \end{bmatrix}.
\end{align}
Such a transformation preserves the bosonic canonical commutation relations of canonical variables~\cite{Arvind}; i.e.,
\begin{align}
\left[\bm{X}_i,\bm{X}_j\right]=\iu\Gamma_{ij},
\end{align}
where $\bm{X}:=(\hat{q}_1,\hat{p}_1,\hat{q}_2,\hat{p}_2,\ldots,\hat{q}_n,\hat{p}_n)$ and the quadrature operators are $\hat{q}_i:=(\hat{a}_i+\hat{a}_i^{\dag})/\sqrt{2}$ and $\hat{p}_i:=(\hat{a}_i-\hat{a}_i^{\dag})/\iu\sqrt{2}$.\\
\begin{figure}
   \centering
   \includegraphics[width=0.9\linewidth]{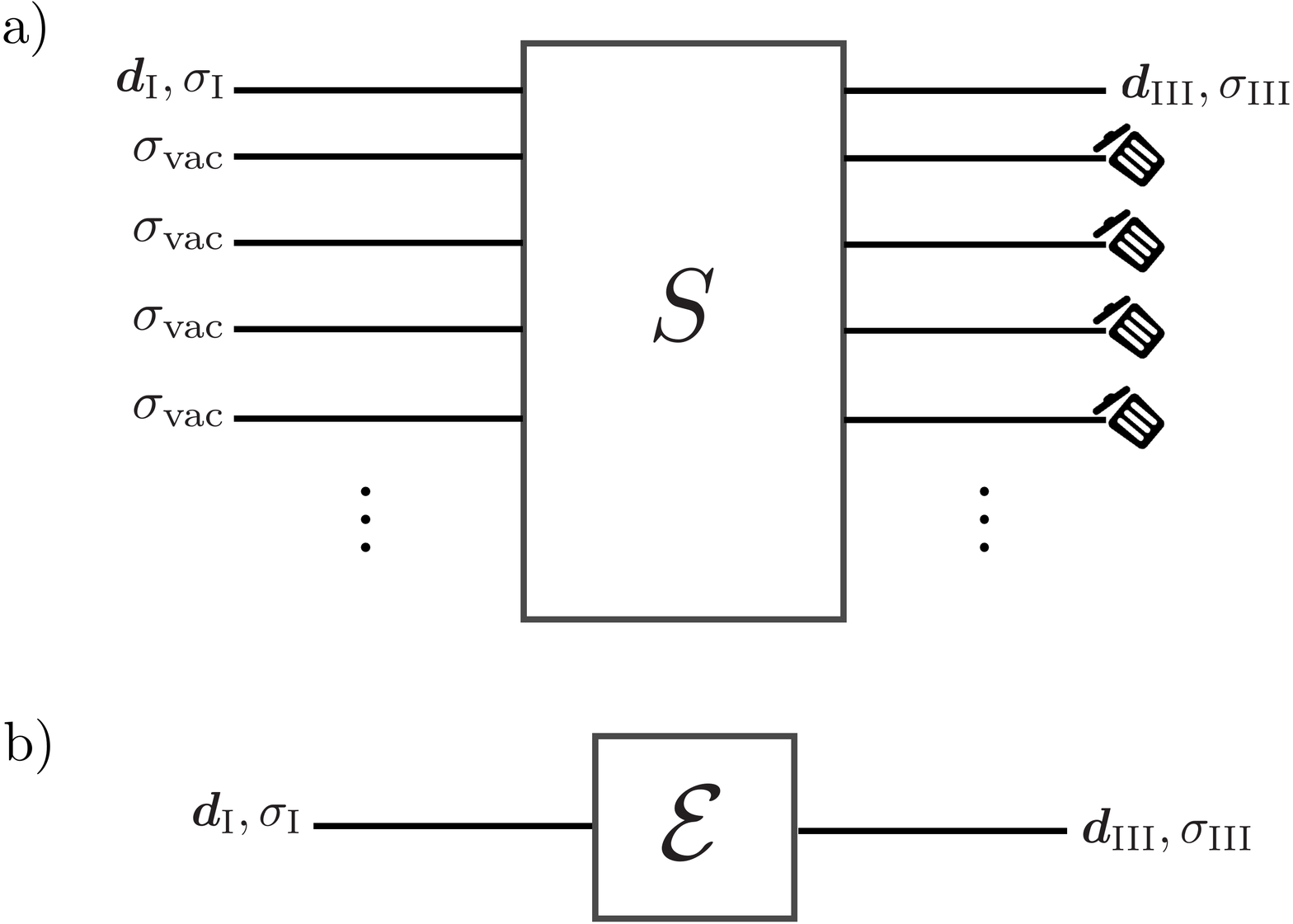} 
   \caption{(a) The BBB is depicted for the case wherein the first mode of the cavity is used to encode and decode quantum information. We assume all the other modes are initially prepared in vacuum and after the BBB (which is represented by the symplectic transformation $S$) the rest of the modes are ignored. (b) The operations performed in part (a) are all Gaussian operations which enables us to express the BBB as a Gaussian channel $\cal E$ acting on the first and second moments as given in Eq.~\eqref{GC}.}
   \label{BBBSymplectic}
\end{figure}
Consider a massless scalar field in a cavity with Dirichlet boundary conditions; i.e., the field vanishes at the two cavity walls.~\footnote{Our analyses can also be performed with Dirac spinor and Maxwell fields and with other boundary conditions such as Neumann boundary conditions~\cite{Friis2013}.} Two complete orthonormal sets of mode functions $\{\phi_n\}$ and $\{\psi_n\}$ can be used to expand the quantum field in regions I and II respectively. These two sets of mode functions and their corresponding ladder operators are related via a Bogoliubov transformation as  
\begin{align}
\psi_j &=\sum_i \tilde{\alpha}_{ij} \phi_i+\tilde{\beta}_{ij}\phi_i^{*},\nn \\
\hat{b}_j&=\sum_i \tilde{\alpha}_{ij}^* \hat{a}_i-\tilde{\beta}_{ij}^* \hat{a}_i^{\dag},		\end{align}
where $\tilde{\alpha}_{ij}:=(\psi_i,\phi_j)$ and $\tilde{\beta}_{ij}:=-(\psi_i,\phi_j^*)$ are the (Minkowski-to-Rindler) Bogoliubov coefficients and $(\cdot 
,\cdot)$ represents the Klein-Gordon inner product~\cite{BirrelDavies}. The transformation  $S_{\text{\tiny I,II}}$ in the quadrature basis is
\begin{align}\label{SM}
S_{\text{\tiny I,II}}&= \begin{bmatrix}
\tilde{M}_{11} & \tilde{M}_{12} & \tilde{M}_{13} & \cdots \\
\tilde{M}_{21} & \tilde{M}_{22} & \tilde{M}_{23} & \cdots \\
\tilde{M}_{31} & \tilde{M}_{32} & \tilde{M}_{33} & \cdots \\
\vdots & \vdots & \vdots & \ddots 
\end{bmatrix}, 
\end{align}
where
\begin{align}
\tilde{M}_{ij}&= \begin{bmatrix}
\Re{\left(\tilde{\alpha}_{ij}-\tilde{\beta}_{ij}\right)} & \Im{\left(\tilde{\alpha}_{ij}+\tilde{\beta}_{ij}\right)} \\
-\Im{\left(\tilde{\alpha}_{ij}-\tilde{\beta}_{ij}\right)} & \Re{\left(\tilde{\alpha}_{ij}+\tilde{\beta}_{ij}\right)} 
\end{bmatrix}.
\end{align}	

We denote the free evolution of the field in region II by $S_{\text{\tiny II}}$ which reads as $S_{\text{\tiny II}}=\bigoplus_{j}G_{j}$, where $G_{j}$ is a rotation in phase space. The transformation from the accelerating frame back to the inertial frame is the inverse of $S_{\text{\tiny I,II}}$; $S_{\text{\tiny II,III}}=S^{-1}_{\text{\tiny I,II}}$. Hence, the full symplectic transformation representing the evolution of the field from region I to region III is 
\begin{align}
S=S^{-1}_{\text{\tiny I,II}}S_{\text{\tiny II}}S_{\text{\tiny I,II}}=\begin{bmatrix}
M_{11} & M_{12} & M_{13} & \cdots \\
M_{21} & M_{22} & M_{23} & \cdots \\
M_{31} & M_{32} & M_{33} & \cdots \\
\vdots & \vdots & \vdots & \ddots 
\end{bmatrix},
\end{align}
where 
\begin{align}
M_{ij}= \begin{bmatrix}
\Re{\left(\alpha_{ij}-\beta_{ij}\right)} & \Im{\left(\alpha_{ij}+\beta_{ij}\right)} \\
-\Im{\left(\alpha_{ij}-\beta_{ij}\right)} & \Re{\left(\alpha_{ij}+\beta_{ij}\right)} 
\end{bmatrix}.
\end{align}	
\begin{figure}
   \centering
   \includegraphics[width=0.9\linewidth]{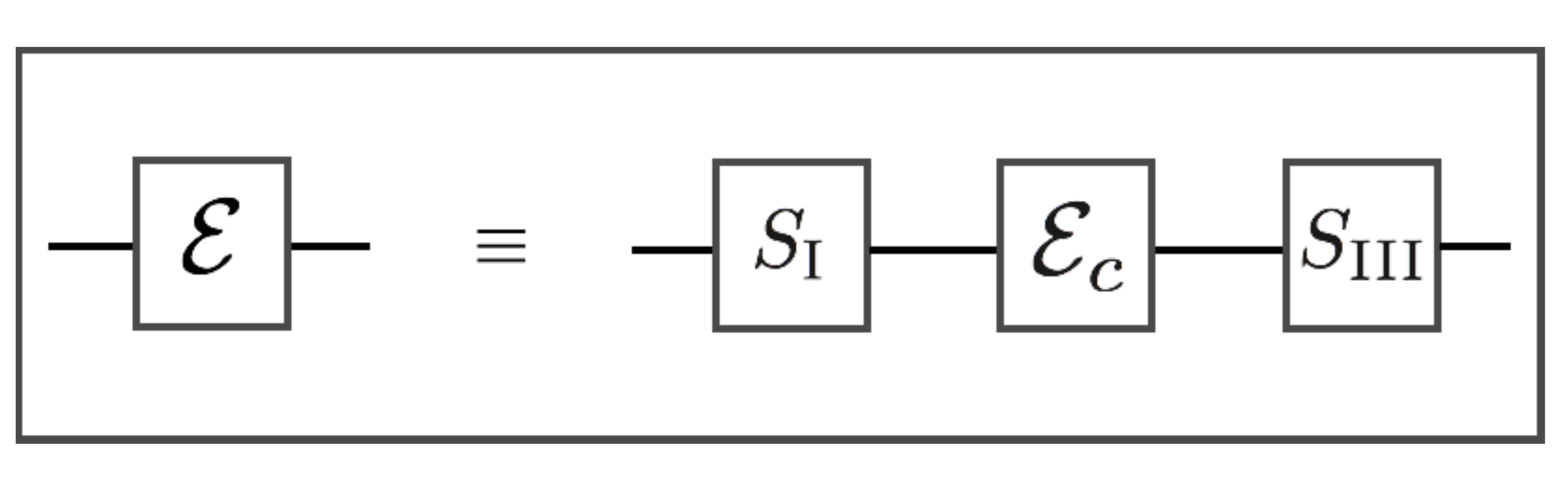} 
   \caption{The canonical form of the BBB Gaussian channel, $\cal E$, which is decomposed into its canonical form, ${\cal E}_c$, up to two symplectic transformations in regions I and III, i.e., $S_{\text{I}}$ and $S_{\text{III}}$.}
   \label{Canonical}
\end{figure}
 The Bogoliubov coefficients are calculated perturbatively in $h=aL$, where $a$ is the proper acceleration at the center of the cavity and $L$ is the cavity length, such that 
\begin{align}\label{PerBogo}
\alpha_{ij}&=\alpha^{(0)}_{ij}+\alpha^{(1)}_{ij}h+\alpha^{(2)}_{ij}h^2+ O(h^3),\nn\\
\beta_{ij}&=\beta^{(0)}_{ij}+\beta^{(1)}_{ij}h+\beta^{(2)}_{ij}h^2+ O(h^3).
\end{align}
Also, from \eqref{Sym}, one observes that the Bogoliubov identity holds; i.e.,
\begin{align}\label{BGidentity}
\sum_{i} \left|\alpha_{ij}\right|^2-\left|\beta_{ij}\right|^2=1.
	\end{align}
These perturbative Bogoliubov coefficients were computed~\cite{Bruschi2012} and, in particular, if $(i+j)$ is even then $\alpha^{(1)}_{ij}=\beta^{(1)}_{ij}=0$. Using the perturbative expansions of the Bogoliubov coefficients \eqref{PerBogo}, for the zero- and second-order terms of \eqref{BGidentity} we get 
\begin{subequations}\label{BGI}
\begin{align}
&\left|\alpha_{ij}^{(0)}\right|^2=1,\label{Phase}\\
&\Re\left({\alpha_{jj}^{(0)}}^{*}\alpha_{jj}^{(2)}\right)+f_{\alpha,k}-f_{\beta,k}=0,
\end{align}
\end{subequations}
where 
\begin{align}
	f_{\alpha,k}&:=\frac{1}{2}\sum_{n\neq k}\left|\alpha_{nk}^{(1)}\right|^2,\nn\\
f_{\beta,k}&:=\frac{1}{2}\sum_{n\neq k}\left|\beta_{nk}^{(1)}\right|^2,
\end{align}
and first-order terms are zero.\footnote{The Bogoliubov identities in the perturbative regime are also obtained using the fact that the change of basis from region I to region III is a unitary operation on cavity modes~\cite{Friis}.}
From \eqref{Phase} we conclude that $\alpha_{ij}^{(0)}=\delta_{ij}\eu^{i \phi_j}$, where
\begin{equation}
	\phi_j:=2\pi j u,\;
	u:= \frac{h \tau}{4L\arctanh(h/2)}.
\end{equation}
Here, $\phi_j$ is a phase that mode $j$ picks up during the accelerating motion of the cavity for the proper time~$\tau$ with respect to the centre of the cavity~\cite{Friis}, i.e., region~II.\\

In this section, we reviewed the BBB for a cavity of size $L$ which accelerates for some time with a fixed proper acceleration $a$ with respect to the centre of the cavity. We employed the symplectic nature of the transformation from the inertial frame to the accelerating frame to write the Bogoliubov identities up to second order. We derived relations~\eqref{BGI} for the perturbative Bogoliubov coefficients which helps us simplify the expressions for  the fidelity of quantum secret sharing in different scenarios.

 %%%%%%%%%%%%%%%%%%%%%%%%%%%%%%%%%%%%%%%%%%%%%%%%%%%%%%%%%%%%%%%%%%%%%%%%%
\section{Methods}\label{Methods}
In this section, we employ the framework of Gaussian quantum information~\cite{Weedbrook2012,Gerardo2014} to write the evolution of the quantum field inside the cavity in a BBB, as depicted in Fig.~\ref{BBB}, as a Gaussian quantum channel. We use this channel, in Section~\ref{Results}, to study the effect of non-inertial motion of the shares on the fidelity of the quantum secret sharing. Moreover, we characterize the canonical form of the channel and show that it is a thermal lossy channel. To this aim, we summerize the framework of Gaussian quantum channels.\\

\begin{figure}
   \centering
\includegraphics[width=0.9\linewidth]{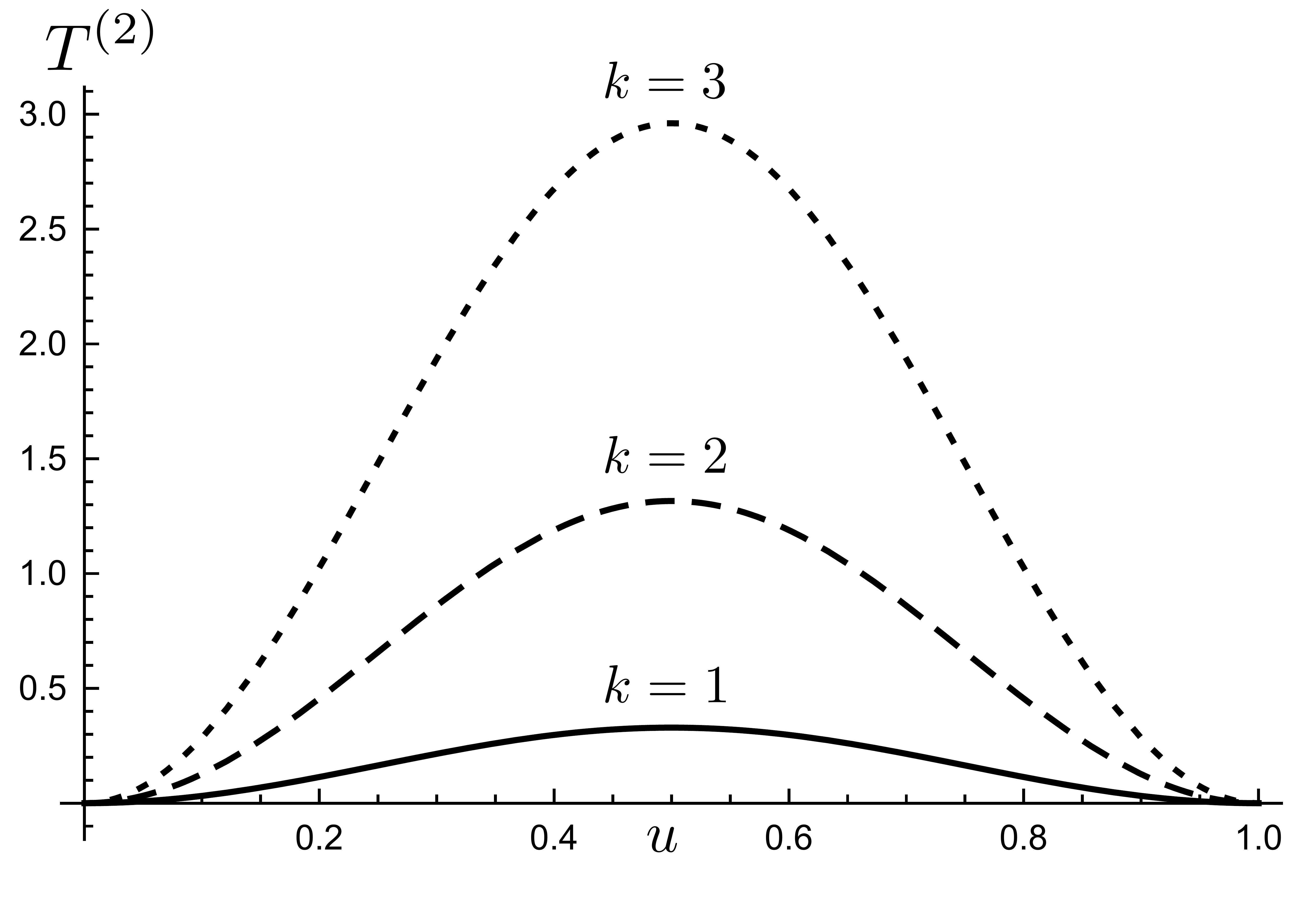} 
   \caption{The second-order coefficient of the transmissivity of the BBB channel, $T^{(2)}$, as a function of $u$ for modes $k=1,2,3$.}
   \label{t2}
\end{figure}
Gaussian states are completely characterized by their first and second moments and, for an $n$-mode Gaussian state, these are
\begin{subequations}\label{GC}
\begin{align}
\bm{d}&=\left(\avg{\bm{X}_1},\avg{\bm{X}_2},\dots,\avg{\bm{X}_n}\right),\\
\sigma_{ij}&=\avg{\bm{X}_i\bm{X}_j+\bm{X}_j\bm{X}_i}-2\avg{\bm{X}_i}\avg{\bm{X}_j}.
\end{align}
\end{subequations}
By definition, Gaussian channels are the subset of quantum channels that transform Gaussian states to Gaussian states. The most general form of a Gaussian channel $\cal E$ is expressed in terms of its action on the first and second moments of the input states as
\begin{subequations}\label{GCH}
\begin{align}
\bm{d}&\xmapsto{\cal E} M \bm{d}+\bm{\alpha},\\
\sigma &\xmapsto[]{\cal E} M \sigma M^T + N,
\end{align}
\end{subequations}
where, for $n$-mode Gaussian channels, $M$ and $N$ are real $2n\times 2n$ matrices, $\bm{\alpha}\in \mathbb{R}^{2n}$ is a displacement vector and $N$ is a symmetric matrix; i.e., $N=N^T$.\\

In Fig.~\ref{BBBSymplectic}, we have depicted the scenario wherein all the modes of the cavity are prepared in the vacuum state except mode $k$, which is prepared in a Gaussian state with first and second moments $\bm{d}_{\text{I}}$ and $\sigma_{\text{I}}$ respectively. First, the initial state of the cavity evolves through the symplectic transformation $S$ and subsequently  all the modes except mode $k$ are traced out. As both the symplectic operation and the tracing operation preserve the Gaussianity of a quantum state, the BBB can be written as a Gaussian channel, which transforms the initial Gaussian state to another one as given in Eq.~\eqref{GC}. Hence, using~\eqref{SM}, the matrices $M$ and $N$ for mode $k$ read as
\begin{subequations}\label{MNNP}
\begin{align}
	M_{kk}&= \begin{bmatrix}
\Re{\left(\alpha_{kk}-\beta_{kk}\right)} & \Im{\left(\alpha_{kk}+\beta_{kk}\right)} \\
-\Im{\left(\alpha_{kk}-\beta_{kk}\right)} & \Re{\left(\alpha_{kk}+\beta_{kk}\right)} 
\end{bmatrix},\\
N_k&=\sum_{n\neq k} M_{nk}M_{nk}^{T}.
\end{align}	
\end{subequations}
\begin{figure}
   \centering
   \includegraphics[width=\linewidth]{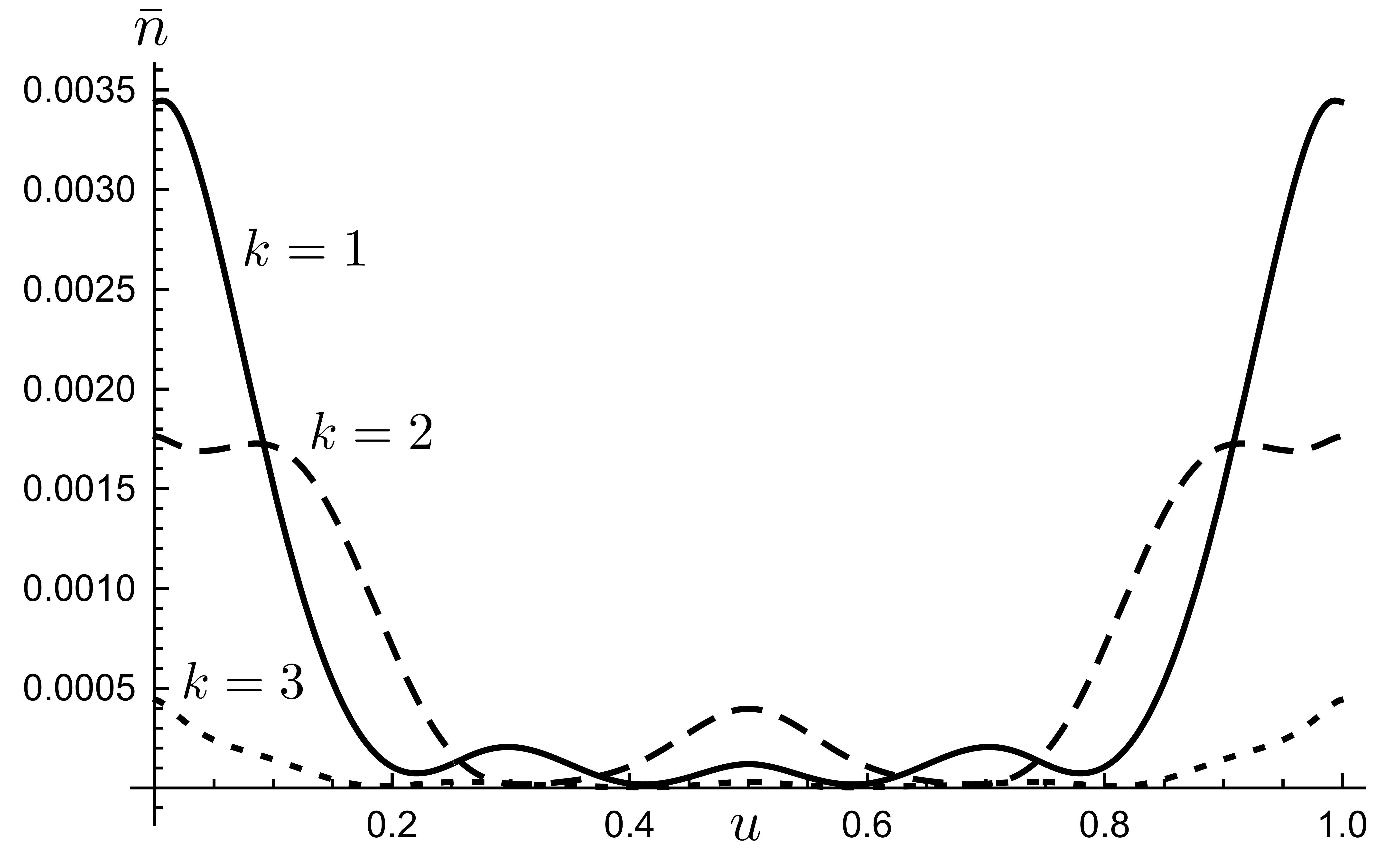} 
   \caption{The average number of thermal particles as a function of $u$ for modes $k=1,2,3$.}
   \label{nbar}
\end{figure}
We are interested in the final quantum state up to third order in $h=aL$, which means that we only need the matrices in Eq.~\eqref{MNNP} up to (but not including) third order in $h$; i.e., 
\begin{align}\label{MNE1}
M_{kk}&=M^{(0)}_{\phi_a}+M^{(2)}_{kk}h^2+ O(h^3),\\
N_k&=N^{(2)}_kh^2+ O(h^3),\nn\\
N^{(2)}_k&=\sum_{n\neq k} M_{nk}^{(1)}{M_{nk}^{(1)}}^{T},\nn\\
M^{(0)}_{\phi_a}&=\begin{bmatrix}
\cos\phi_a & \sin\phi_a \nn\\
-\sin\phi_a & \cos\phi_a
\end{bmatrix}, \\
M^{(0)}_{nk}&=M^{(1)}_{kk}=0 \,\,(n\neq k),\nn \\
M^{(i)}_{nk}&=\begin{bmatrix}
\Re{\left(\alpha^{(i)}_{nk}-\beta^{(i)}_{nk}\right)} & \Im{\left(\alpha^{(i)}_{nk}+\beta^{(i)}_{nk}\right)} \\
-\Im{\left(\alpha^{(i)}_{nk}-\beta^{(i)}_{nk}\right)} & \Re{\left(\alpha^{(i)}_{nk}+\beta^{(i)}_{nk}\right)} 
\end{bmatrix},\nn
\end{align}
where in the last matrix $i=1,2$. We emphasize that as we are estimating the effect of the Gaussian channel up to third order in $h$, the term $M^{(2)}\sigma_\text{I}{M^{(2)}}^{T}$ is to be ignored.\\

As was pointed out by Holevo~\cite{Holevo2007}, any Gaussian quantum channel can be decomposed into its canonical form and two Gaussian unitary operators; one on the input and one on the output. This means that we can decompose the Gaussian channel for the BBB as shown in Fig.~\ref{Canonical}.
Here, $S_{\text{\tiny I}}$ and $S_{\text{\tiny III}}$ are two symplectic transformations in the region I and III, which correspond to the two Gaussian unitary operators. We use $M_c$ and $N_c$ for the canonical form of the channel, ${\cal E}_c$, as opposed to the channel $\cal E$ for which we have used $M$ and $N$.\\

\begin{figure}
\centering
\includegraphics[width=0.85\linewidth]{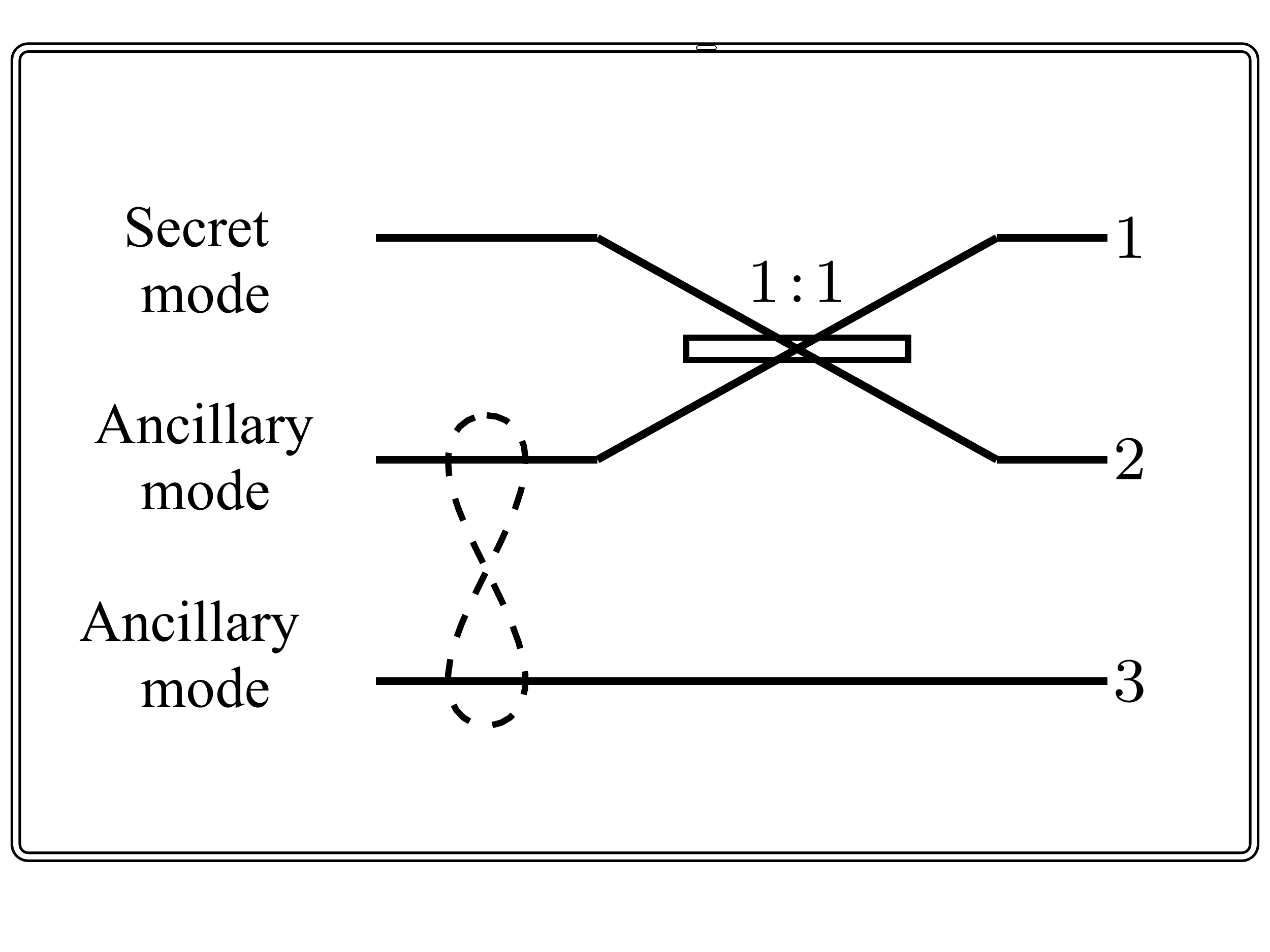}
\caption{The encoding circuit for continuous-variable $(2, 3)$~-~threshold quantum secret sharing. The ``8'' shaped sympbol represents a two-mode squeezed-vacuum state. The upper two modes are combined on a balanced beam splitter. The three outputs are three quantum shares, denoted by mode~$1$, mode~$2$, and mode~$3$.}\label{encoding}
\end{figure}

In transforming a Gaussian quantum channel $\cal E$ to its canonical form ${\cal E}_c$, some properties of the channel remain invariant (up to symplectic transformations $S_{\text{\tiny I}}$ and $S_{\text{\tiny III}}$). The first invariant is $r:=\min\left[\text{rank}(M),\text{rank}(N)\right]$ which, for a single-mode channel, can take the possible values $r=0,1,2$ and in our case we have $r=2$. The second invariant is the \textsl{transmissivity} of the channel,\begin{align}
	T =\det M=1-T^{(2)}h^2+ O(h^3),
	\end{align}
where
\begin{align}
T^{(2)}:=2\left( f_{\alpha,k}-f_{\beta,k}\right).\nn
\end{align}
 Note that, as $T^{(2)}$ increases, the transmissivity decreases. In Fig.~\ref{t2}, we plot $T^{(2)}$ as a function of $u$, where we observe that by increasing the mode number $k$ the transmissivity of the channel decreases. Here we choose to plot all the quantities in terms of $u$, as the Bogoliubov coefficients for a BBB are periodic in $u$ with the period of $1$. \\
\begin{figure}
\centering
\includegraphics[width=1.05\linewidth]{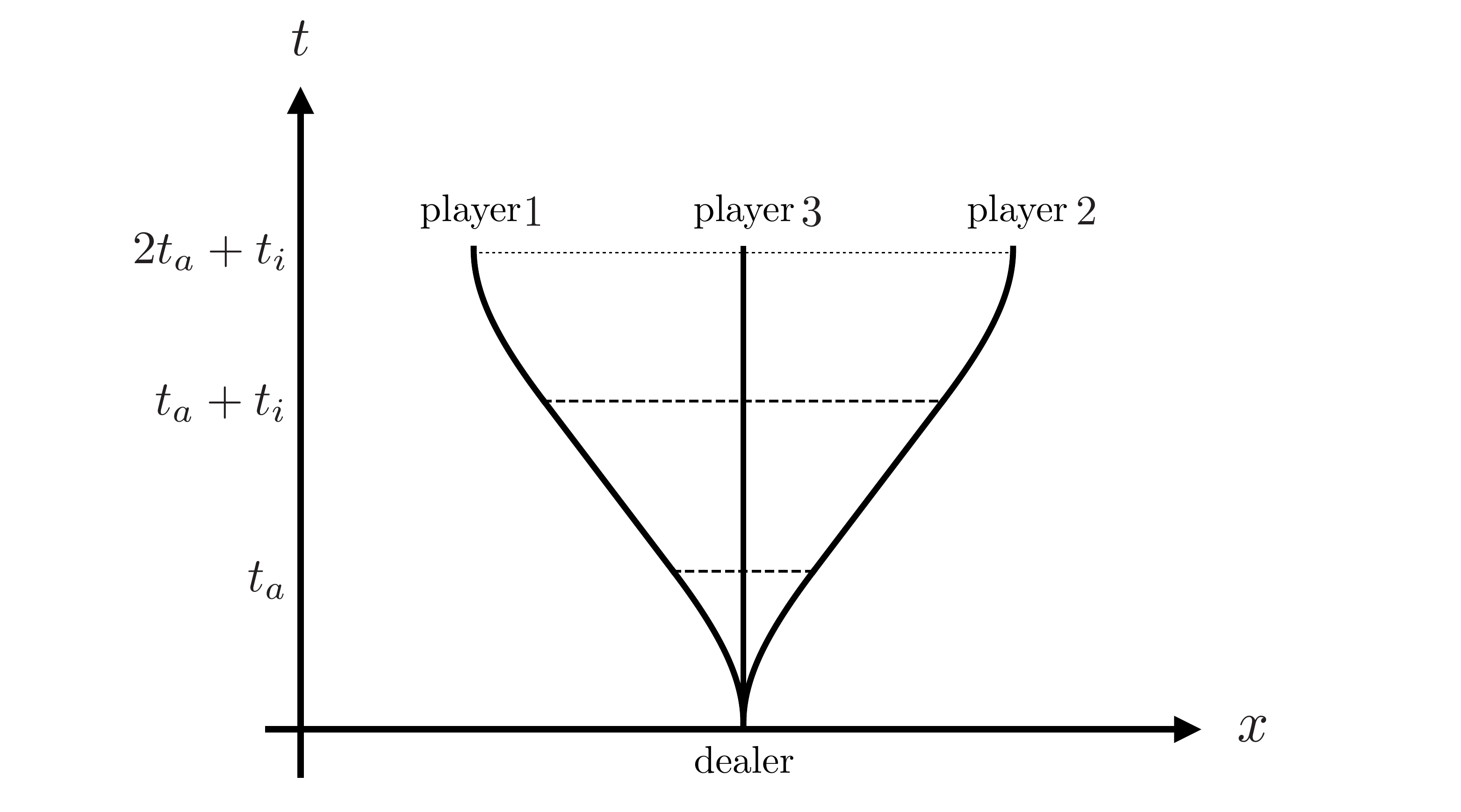}
\caption{The worldlines of the quantum shares during distribution. From $t=0$ to $t=t_a$, the two cavities, represented by the furthest left and the furthest right worldlines, accelerate with the proper acceleration $a$ in two opposite directions. From $t=t_a$ to $t=t_a+t_i$, they move with constant velocities. From $t=t_a+t_i$ to $t=2t_a+t_i$, the two cavities decelerate with the proper acceleration $a$ and  become stationary. The cavity represented by the middle world line remains static.}
\label{fig:distribution}
\end{figure}

The final invariant is thermal number $\bar{n}$ associated to the canonical form of the quantum channel \cal{E}. We calculate the leading order term of $\bar{n}$, which is
\begin{align}
\bar{n}&:=\frac{\sqrt{\det{N}}}{2|1-T|}-\frac{1}{2}=\frac{\sqrt{\left(f_{\alpha,k}+f_{\beta,k}\right)^2-4\left|g_{\alpha\beta,k}\right|^2}}{2\left(f_{\alpha,k}-f_{\beta,k}\right)}-\frac{1}{2}, 
\end{align}
where $g_{\alpha\beta,k}:=\sum_{n\neq k}\alpha^{(1)}_{nk}\beta^{(1)}_{nk}$. In Fig.~\ref{nbar}, we plot this quantity as a function of $u$.\\

The main advantage in working with the canonical form of the BBB channel is that we can completely charactrize it. For the symplectic invariants we find $T\in(0,1)$ and $r=2$, from which we can conclude that the canonical form of the BBB channel is a thermal lossy Gaussian channel. The channel is lossy due to the fact that its transmissivity is smaller than one; i.e., $T<1$. Furthermore, from this analyses, we conclude that the quantum channel ${\cal E}_c$ can be simulated by interacting mode $k$ of the cavity and a thermal state with mean photon number $\bar{n}$ via a beam splitter of transmittance $T$.\\

In this section, we employed the framework of Gaussian channels to find matrices $M$ and $N$ in \eqref{GCH} for a BBB. From this point on, we use them to include the effect of relativity on the quantum field inside a cavity while the cavity moves non-inertially. Moreover, we computed the channel invariants, transmissivity and the average number of thermal particles, which enabled us to  identify the BBB as a thermal lossy channel.    

\section{The relativistic protocol}\label{Results}
In this section, we present  the relativistic variant of (2,3)-threshold continuous-variable quantum secret sharing. We first include the effect of acceleration on the distribution of quantum shares and then we consider different possible collaboration scenarios between the players. In each case, we show that the fidelity of quantum secret sharing is reduced, except for a thermal state, when compared to the non-relativistic scenarios.\\

%The covariance matrix of this three-mode Gaussian state is
%\begin{align}
%\sigma_{\text{en}}=\left(B_\frac{1}{2}\oplus \mathbb{I}\right) (\sigma\oplus \sigma_{\text{\tiny TMSS}})\left(B_\frac{1}{2}\oplus \mathbb{I}\right)^T,
%\end{align}
%where 
%\begin{align}
%B_\eta=\begin{bmatrix}
  %\sqrt{\eta}\, \mathbb{I}   &  \sqrt{1-\eta}\,\mathbb{I} \\
  %-\sqrt{1-\eta}\, \mathbb{I}  &  \sqrt{\eta}\, \mathbb{I}  \\  
%\end{bmatrix}
%\end{align} 
%represents the symplectic matrix of a beam splitter with reflectivity $\eta$, and $\mathbb{I}$ represents the identity operation on the third mode.\\  

\subsection{Distribution of quantum shares} 
In our case, modes 1, 2, and 3 are three  quantum Gaussian shares and each mode corresponds to a mode in a cavity. The Gaussian state of each quantum share occupies one mode inside each cavity, and the other modes inside each cavity are all in vacuum states. Fig.~\ref{encoding} shows the encoding protocol of a $(2, 3)$-threshold quantum secret sharing scheme as proposed by Tyc and
Sanders~\cite{Tyc1}. The dealer encodes the quantum secret into a three-mode Gaussian state, i.e., modes 1, 2, and 3. He prepares modes 2 and 3 in a two-mode squeezed-vacuum state, with the squeezing parameter $s$. Then, he combines modes~1 (the quantum secret) and~2 on a balanced beam splitter. The two output beams of the beam splitter, together with the third beam encoding the other half of the two-mode squeezed state, are the three quantum shares, which are distributed to the three players.\\
\begin{figure}
\centering
\includegraphics[width=1.05\linewidth]{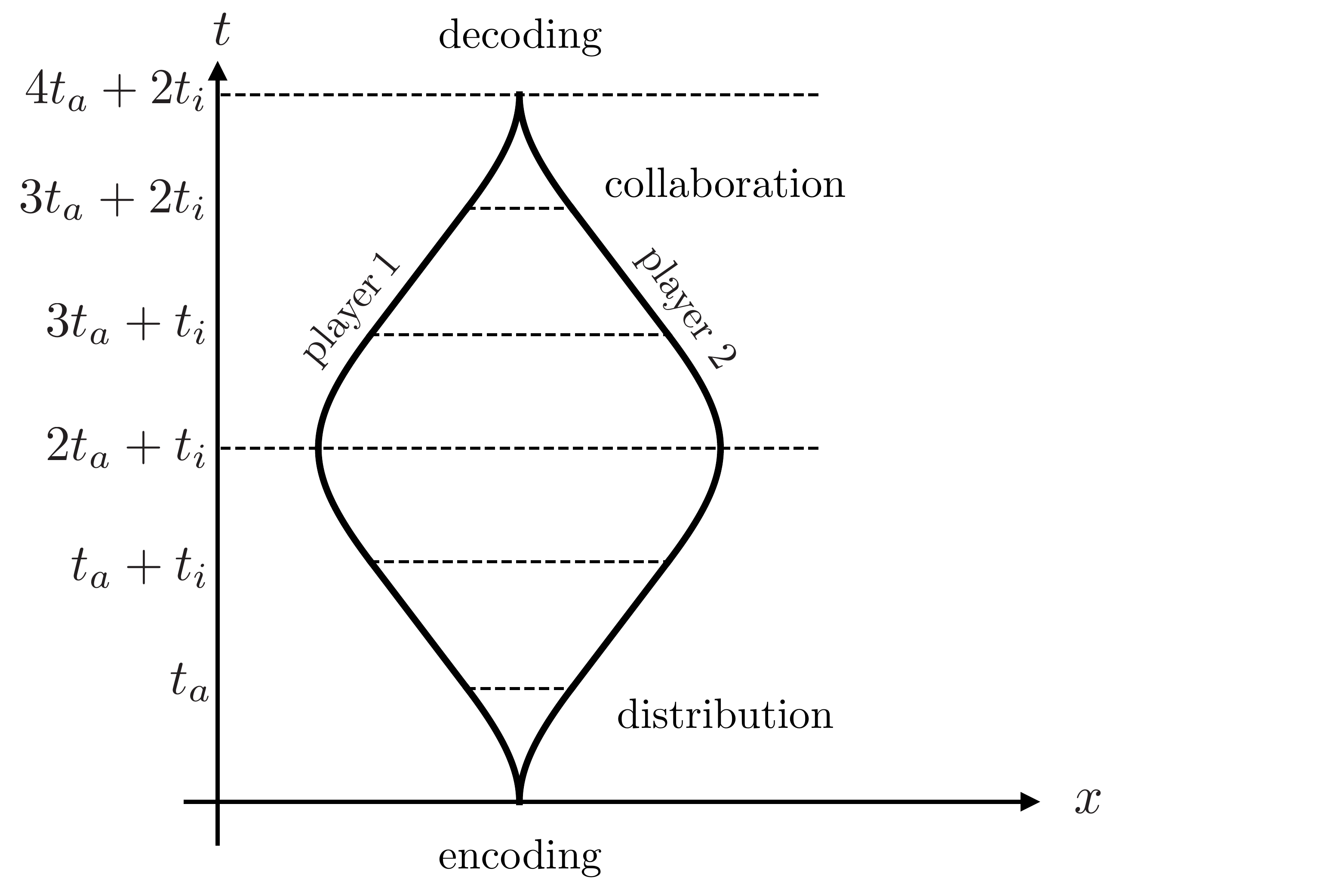}
\caption{The two curves represent two worldlines in spacetime. Each worldline is the trajectory of one cavity carrying a quantum share. From $t=2t_a+t_i$ to $t=3t_a+t_i$, the two cavities accelerate with proper acceleration $a$ towards each other. From $t=3t_a+t_i$ to $t=3t_a+2t_i$, the two cavities are moving with constant velocity. From $t=3t_a+2t_i$ to $t=4t_a+2t_i$, the two cavities decelerate with proper acceleration $a$ to arrive at the same spacetime point.}
\label{fig:collaboration12}
\end{figure}

After encoding, the dealer distributes the three cavities to the three players. Fig.~\ref{fig:distribution} shows the distribution of the quantum shares. The three players are located at different spacetime points. One player (Player~3) is at the same spatial position as the dealer and the other two players (Players~1 and~2) have the same distance to the dealer\footnote{To simplify the calculations, we have chosen the symmetric configuration of the players and the dealer, Fig.~\ref{fig:distribution}, which suffices to study the relativistic effects in the distribution stage.}. The dealer and three players are all static, so they share an inertial frame. As depicted in Fig.~\ref{fig:distribution}, Cavities~1 and~3 inevitably need to be accelerated and then decelerated to reach the spacetime regions of Players~1 and~3 respectively. Using the quantum channel derived in Sec.~\ref{Methods}, we consider the effect of such a non-uniform motion of the cavities on the quantum share, which is encoded in a single mode of each cavity. In this scenario, Cavity~3 remains static during the whole distribution.\\  
\subsection{Players' collaboration}
After the quantum shares are distributed between the three players, two of them need to collaborate to decode the quantum secret. Three different scenarios are possible; Players 1 and 3, 2 and 3, or 1 and 2 can constitute the subset of collaborating players. The effect of acceleration on the fidelity of quantum secret sharing in the latter two cases is the same (due to the present symmetry), and we only consider the scenario wherein Players 2 and 3 collaborate.\\
\begin{figure}
\centering
\includegraphics[width=\linewidth]{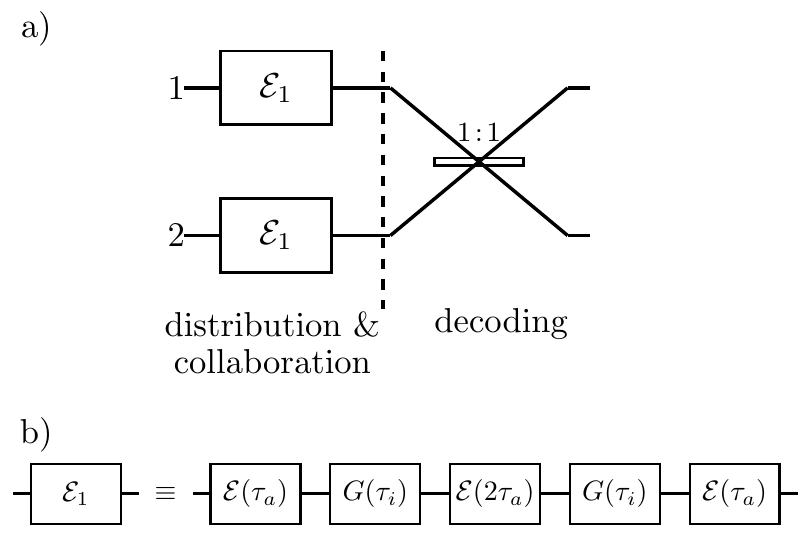}
\caption{(a) The thermal lossy channel $\mathcal E_1$ is the Gaussian channel that represents the total evolution of the first and the second quantum shares during the distribution and the collaboration stage. Then the quantum secret is decoded using a balanced beam splitter. (b) $\mathcal E_1$ is a single-mode Gaussian channel composed of five Gaussian channels in series. $\mathcal{E}(\tau_a)$ is the Gaussian channel for a BBB during the proper time $\tau_a$ and $G(\tau_i)$ represents the Gaussian channel of the free evolution in an inertial frame with proper time $\tau_i$.}
\label{fig:collaborationcircuit12}
\end{figure}

\subsubsection{Collaboration between Players~1 and~2 }
First, we consider the case wherein Players 1 and 2 are collaborating. To decode the quantum secret, their cavities are transported to the same spacetime point as shown in Fig.~\ref{fig:collaboration12}. After the two cavities are at the same region, the quantum secret is decoded by beam splitting the two modes that were employed to encode the quantum secret.  From $t=2t_a+t_i$ to $t=4t_a+2t_i$, each mode of the two-mode Gaussian state goes through the same single-mode Gaussian channel $\mathcal E_1$ as shown in Fig.~\ref{fig:collaborationcircuit12}.\\ 

The Gaussian quantum channel $\mathcal E_1$ is composed of five Gaussian channels in series (See Fig.~9(b)). The channel $\mathcal E (\tau_a)$ corresponds to uniformly accelerated motion of the cavity to the left (or to the right) during the proper time $\tau_a$, while the channel $\mathcal E (2\tau_a)$ represents the cavity moving with constant proper acceleration to the right (or to the left) during the proper time $2\tau_a$. Also, the quantum channel $G(\tau_i)$ corresponds to the inertial movement of the cavity with constant velocity for the proper time $\tau_i$. Using ~\eqref{MNE1}, the first and second moments of the $k$-th mode of the cavity, up to third order in $h$, are transformed as 
\begin{subequations}
\begin{align}
\bm{d} \xmapsto{\mathcal E (\tau_a)} &\, \left(M^{(0)}_{\phi_a}+M^{(2)}_{kk} h^2\right) \bm{d},\\
\sigma \xmapsto{\mathcal E (\tau_a)} &\, M^{(0)}_{\phi_a}\sigma M^{(0)T}_{\phi_a}\nn\\
&+\left(M^{(0)}_{\phi_a}\sigma M^{(2)T}_{kk} +M^{(2)}_{kk}\sigma M^{(0)T}_{\phi_a}\right)h^2+N^{(2)}_{k}h^2,
\end{align}
\end{subequations}
where $M^{(0)}_{\phi_a}$, $M^{(2)}_{kk}$, and $N^{(2)}_{k}$ are given in \eqref{MNE1}.\\

The Gaussian channel $G(\tau_i)$ represents the free evolution of the Gaussian state during the inertial movement of the cavity in proper time $\tau_i$,
\begin{subequations}
\begin{align}
\bm{d} \xmapsto {G (\tau_i)}  &\, M^{(0)}_{\phi_i} \bm{d},\\
\sigma \xmapsto{G (\tau_i)} & \,M^{(0)}_{\phi_i}\sigma M^{(0)T}_{\phi_i},
\end{align}
\end{subequations}
where 
\begin{align}
M^{(0)}_{\phi_i} =
\begin{bmatrix}
    \cos\phi_i &  -\sin\phi_i  \\
   \sin\phi_i  &  \cos\phi_i  
\end{bmatrix}\nn,
\end{align}
and $
\phi_i=\frac{k\pi \tau_i}{L}
$ is the phase accumulated during the free evolution from $t=t_a$ to $t=t_a+t_i$, and from $t=3t_a+t_i$ to $t=3t_a+2t_i$. To simplify the later calculations, we suppose the phase shift during the inertial movement is $\phi_i=\pi-2\phi_a$.\\

Therefore, we can express the collaboration between Players 1 and 3, shown in Fig.~\ref{fig:collaboration12}, as the Gaussian channel $\mathcal E_1$, 
\begin{align}
\mathcal E_1:= \mathcal E(\tau_a)\circ G(\tau_i) \circ \mathcal E(2\tau_a)\circ G(\tau_i)\circ\mathcal E(\tau_a). 
\end{align}
\begin{figure}
\centering
\includegraphics[width=1\linewidth]{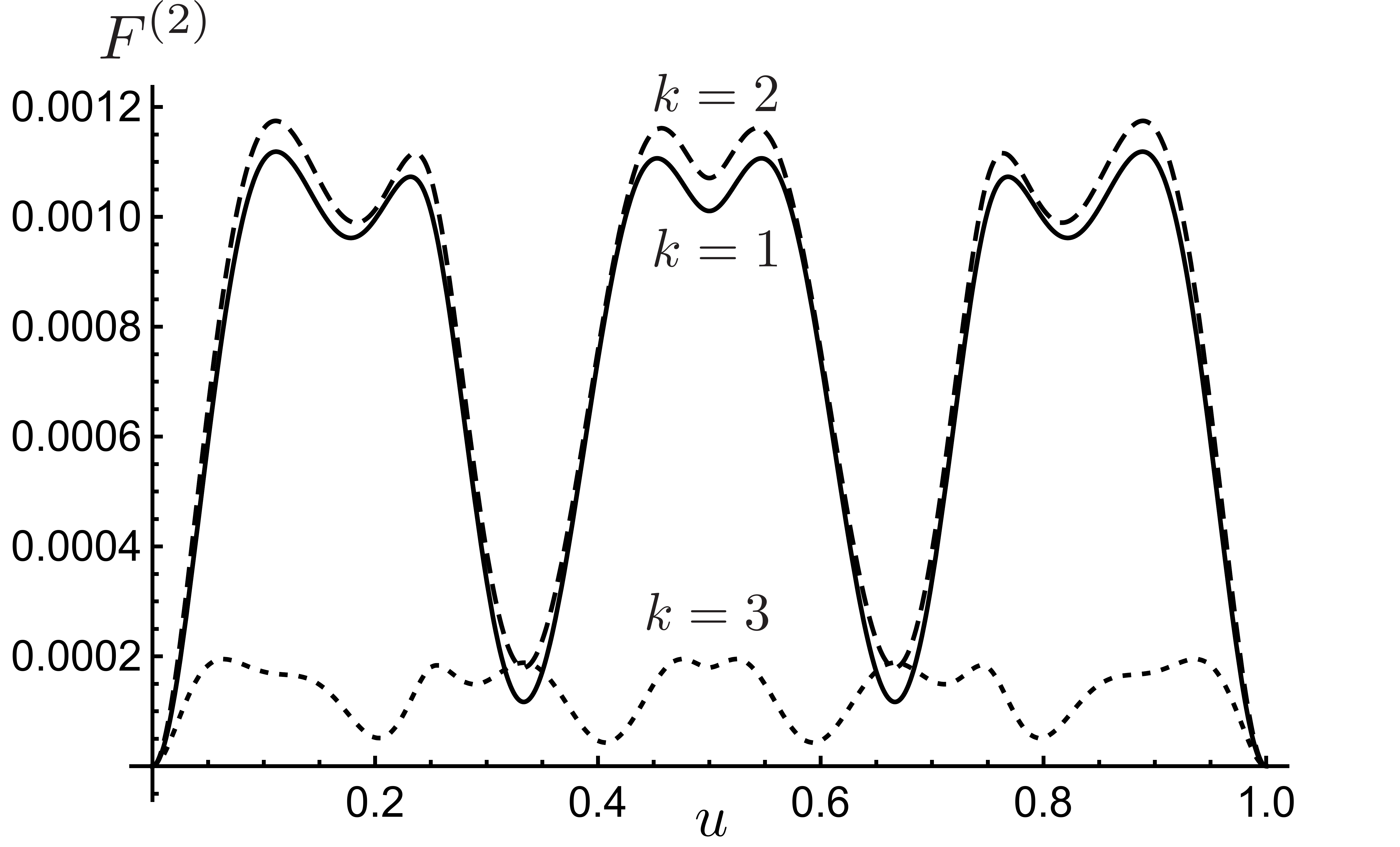}
\caption{$F^{(2)}$ as a function of $u$ for modes $k=1$ (solid), $2$ (dashed), and $3$ (dotted) when the secret Gaussian state is a coherent state.}
\label{fig:coherentFidelity12}
\end{figure}
If the secret Gaussian state is a coherent state and the free evolution is ignored; i.e., $M^{(0)}=\mathbb{I}$,  the Gaussian transformation of the channel $\mathcal E_1$ for the mode $k$ is
\begin{subequations}
\begin{align}
\bm{d}\xmapsto{\mathcal E_1} & \,\bm{d}+(2M_{kk, \tau_a}^{(2)} +M_{kk, 2\tau_a}^{(2)})h^2 \bm{d},\\
 \notag
\mathbb{I} \xmapsto{\mathcal E_1} &\, \mathbb{I}+\Big(2M_{kk, \tau_a}^{(2)}+ 2M_{kk, \tau_a}^{(2)T}+M_{kk, 2\tau_a}^{(2)} +  M_{kk, 2\tau_a}^{(2)T} \\
&\hspace{0.6cm} +2 N_{k, \tau_a}^{(2)} + N_{k, 2\tau_a}^{(2)}\Big) h^2,
\end{align}
\end{subequations}
where $M_{kk, \tau_a}^{(2)}$ and $N_{k, \tau_a}^{(2)} $  are in terms of proper time $\tau_a$.\\
 \begin{figure}
\begin{center}
\includegraphics[width=1\linewidth]{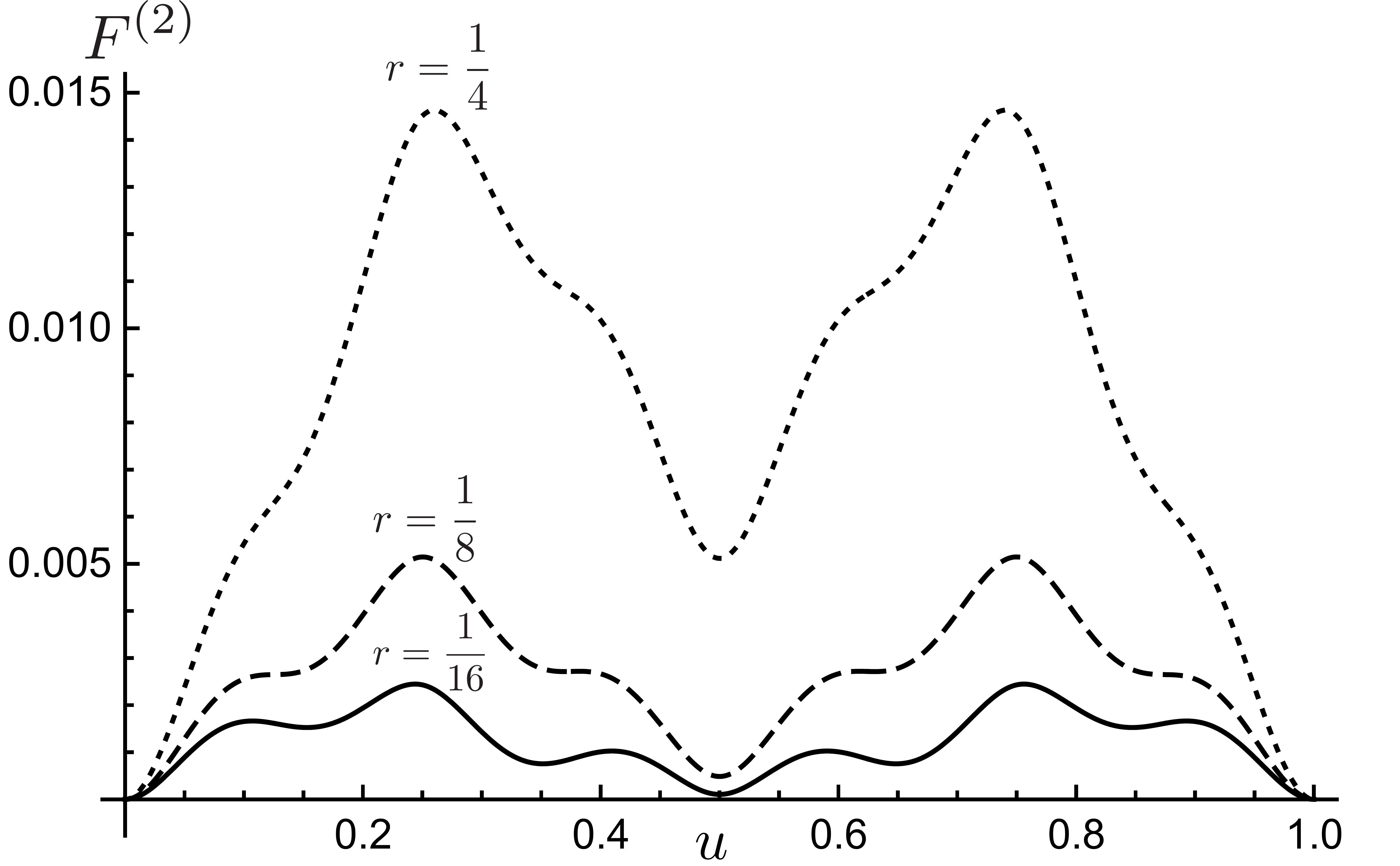}
\caption{$F^{(2)}$ as a function of $u$ for the ground mode $(k=1)$, when the secret Gaussian state is a squeezed-vacuum state for squeezing parameters $r=\frac{1}{16}$ (solid), $\frac{1}{8}$ (dashed), and $\frac{1}{4}$ (dotted).}
\label{fig:squeezingFidelity12}
\end{center}
\end{figure}
 
After the two cavities arrive at the same spacetime region, the two Gaussian quantum shares are combined using a balanced beam splitter, as shown in Fig.~\ref{fig:collaborationcircuit12}. The decoded Gaussian quantum secret is not a pure state anymore due to the effect of acceleration during distribution and collaboration. For a coherent state as the encoded secret Gaussian state, we calculate the fidelity of the quantum secret sharing~\cite{Lance1,Lance2}
\begin{align}\label{coherentfid13} 
F=1-2 \left(2f_{\beta, k, 2u} + f_{\beta, k, u} \right)h^2 + O(h^3).
\end{align}
Interestingly, from ~\eqref{coherentfid13}, we conclude that the fidelity for a coherent state is independent of the initial mean photon number of the quantum secret. In other words, the fidelity of a coherent state is the same as the fidelity of the vacuum state.\\

In Fig.~\ref{fig:squeezingFidelity12}, we have plotted the second-order coefficient of the fidelity, $ F^{(2)}$, for a squeezed-vacuum quantum secret; i.e., $F=1-F^{(2)}h^2$. The figure shows that the fidelity decreases as the squeezing parameter $r$ increases, i.e., as the mean photon number in the secret increases. This is in contrast to the case where the secret state is a coherent state.
\begin{figure}
\begin{center}
\includegraphics[width=1.05\linewidth]{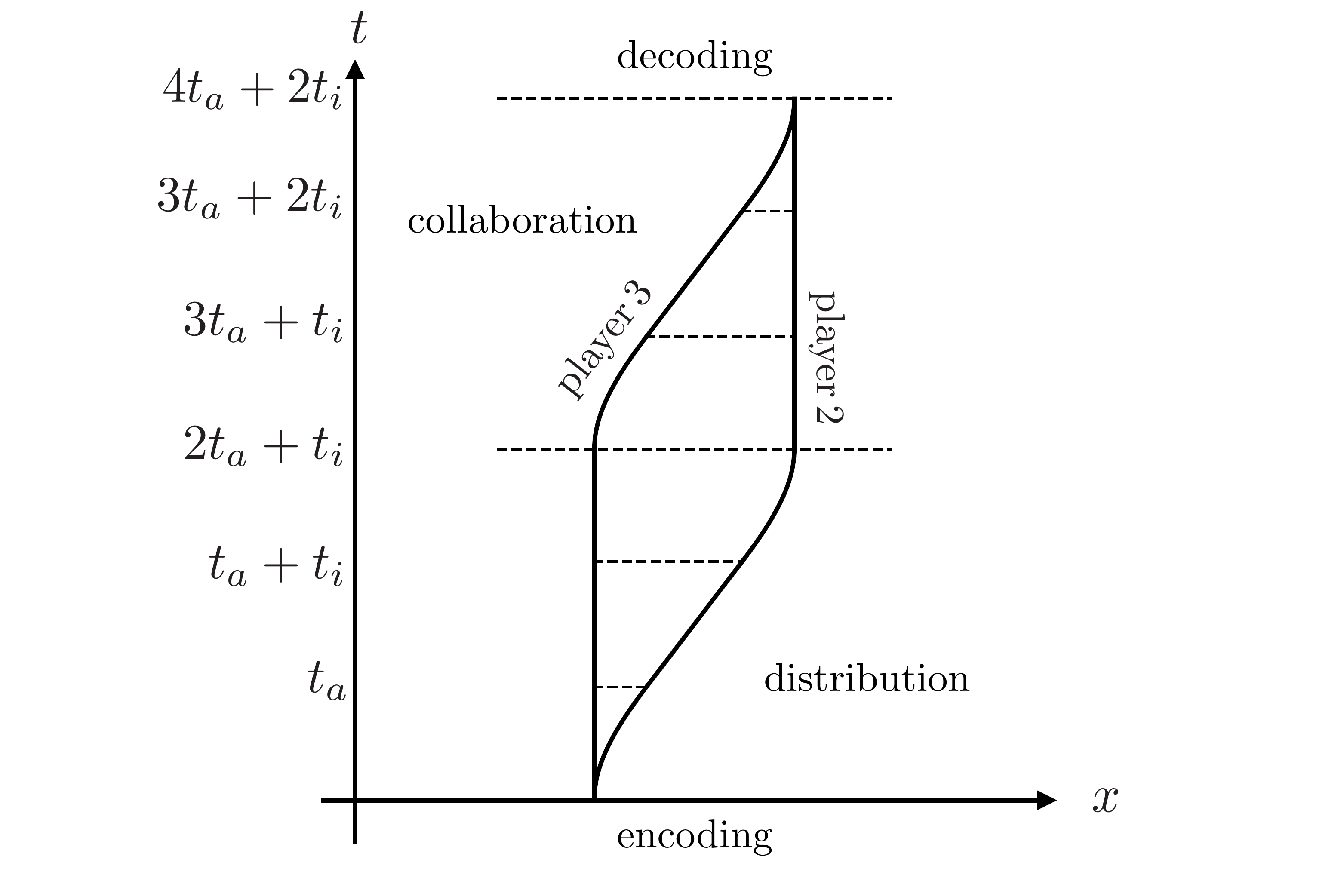}
\caption{The two curves represent two worldlines in spacetime. The left worldline is the trajectory of the cavity carrying the third quantum share and the right worldline is the trajectory of the cavity carrying the second quantum share. From $t=2t_a+t_i$ to $t=4t_a+2t_i$, the third cavity remains static. From $t=2t_a+t_i$ to $t=3t_a+t_i$, the second cavity accelerates with proper acceleration $a$. From $t=3t_a+t_i$ to $t=3t_a+2t_i$, it moves with constant velocity and from $t=3t_a+2t_i$ to $t=4t_a+2t_i$, decelerates with proper acceleration $a$.}
\label{fig:collaboration23}
\end{center}
\end{figure}
\subsubsection{Collaboration between Players~2 and~3 }
\begin{figure}
\begin{center}
\includegraphics[width=1\linewidth]{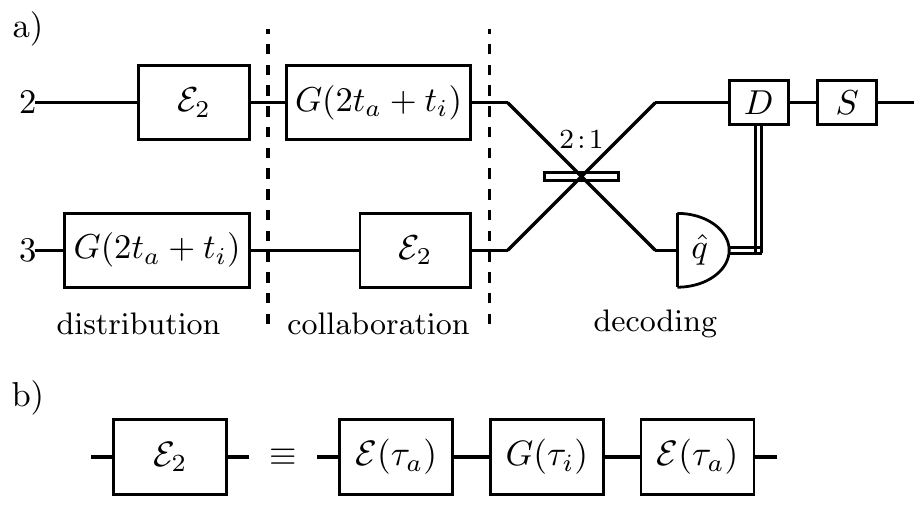}
\caption{(a) The decoding circuit for the case wherein players~$2$ and~$3$ collaborate. $\mathcal E_2$ is a Gaussian thermal lossy channel. $G(2t_a+t_i)$ is the free evolution in the inertial frame.  First, the two modes are combined on a beam splitter with reflectivity $2/3$. Then the quadrature $\hat{q}$ of the second output mode is measured and a displacement operation controlled by the measurement outcome and a squeezing operation are applied on the first output mode. (b) $\mathcal E_2$ is a single-mode Gaussian channel composed of three Gaussian channels in series.}
\label{fig:collaborationcircuit23}
\end{center}
\end{figure}

The second collaboration scenario we consider is the case wherein Players~2 and~3 collaborate to reconstruct the secret quantum state\footnote{We emphasize that the collaboration between Players~1 and~2 results in the same results for the fidelity of the quantum secret sharing, which is simply due to the symmetry in the configuration of the players.}. Similar to the previous case, the quantum shares of Players~2 and~3 are first transported to the same spacetime region. Fig.~\ref{fig:collaboration23} shows the trajectories of the two corresponding cavities in this scenario. Note that the trajectory of the second cavity during the collaboration stage is the same as the trajectory of the third cavity during the distribution stage of the protocol. As Fig.~\ref{fig:collaborationcircuit23} shows, the second quantum share goes through the channel $\mathcal E_2$ during distribution, while it goes through the channel $G(2t_a+t_i)$ during collaboration. The third quantum share first goes through the channel $G(2t_a+t_i)$ when the shares are being distributed and then is affected by the channel $\mathcal E_2$, which represents the effect of acceleration on this quantum share during collaboration.\\

\begin{figure}
\centering
\includegraphics[width=1\linewidth]{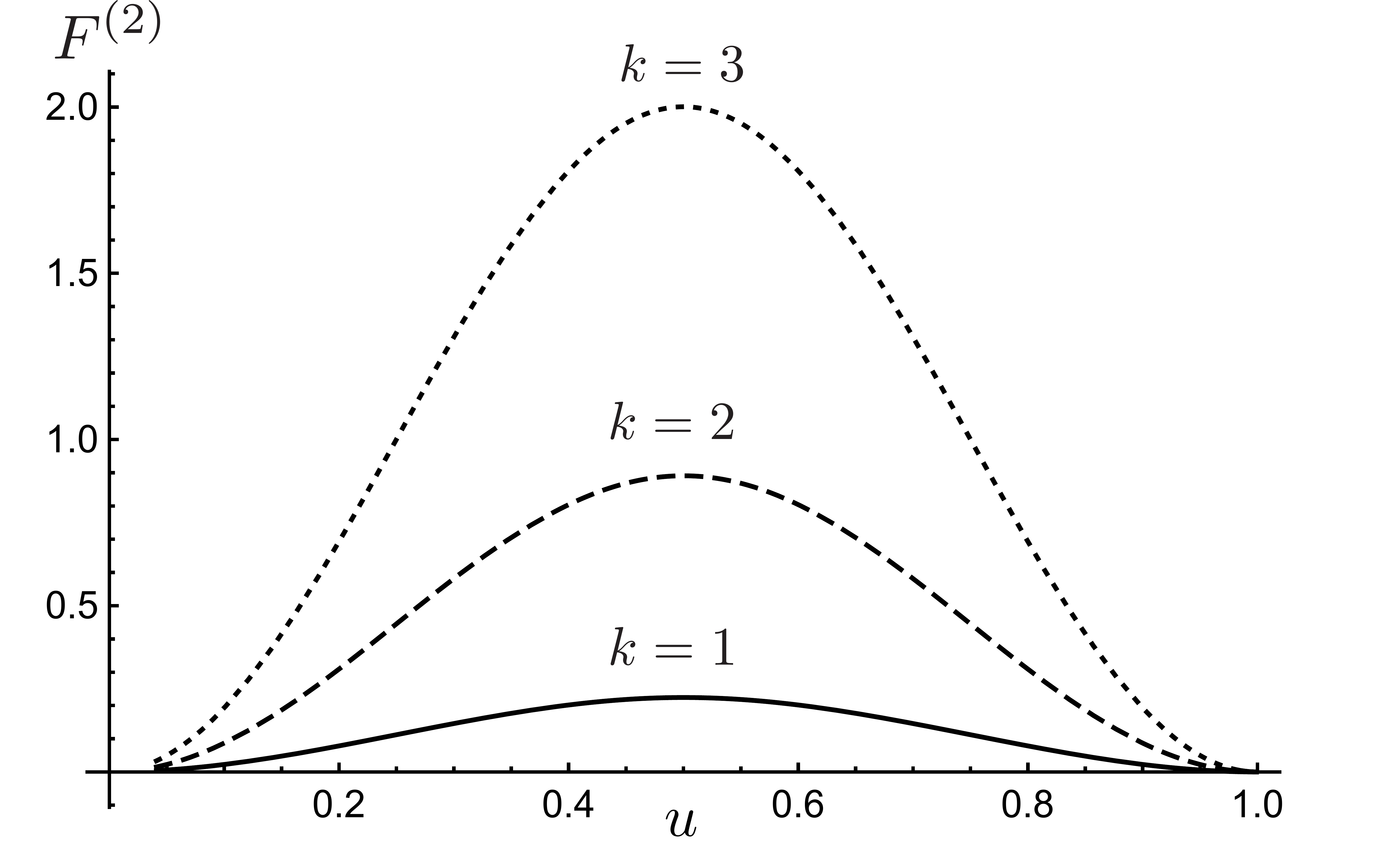}
\caption{$F^{(2)}$ as a function of $u$ for modes $k=1$ (solid), $2$ (dashed), and $3$ (dootted) when the two-mode squeezing parameter $s=1$.}
\label{fig:coherentFidelity23(2)}
\end{figure}

As shown in Fig.~\ref{fig:collaborationcircuit23}, the quantum Gaussian channel $\mathcal E_2$ is a combination of three quantum Gaussian channels, one of which is merely a phase rotation, i.e., $G(\tau_i)$. Assuming the input state of the Gaussian channel $\mathcal E_2$ is a coherent state and the free evolution is ignored, then the transformation of the first and second moments due to the channel $\mathcal E_2$ can be written as 
\begin{subequations}
\begin{align}
\bm{d} \xmapsto{\mathcal E_2} & \,\left(\mathbb{I} + 2M_{kk}^{(2)}\, h^2\right) \bm{d},\\
\mathbb{I} \xmapsto{\mathcal E_2}  &\,\mathbb{I} +
2\left( M_{kk}^{(2)} + M_{kk}^{(2)T} + N_{k}^{(2)}  \right) h^2.
\end{align}
\end{subequations}
After the second and the third quantum shares reach the same spacetime region, the decoding of the quantum secret begins. For decoding, we employ the procedure introduced in~\cite{Lance1,Lance2}. The optical decoding circuit is shown in Fig.~\ref{fig:collaborationcircuit23}, which is applied to reconstruct the secret quantum Gaussian state. We calculate the fidelity of quantum secret sharing in this case up to third order; i.e., 
\begin{align}
F=F^{(0)}-F^{(2)}\,h^2+O(h^3), 	
\end{align}
where $F^{(0)}$ and $F^{(2)}$ are 
\begin{align}
F^{(0)}&=\frac{1}{1+\eu^{-s}},\nn\\
F^{(2)}&=  \frac{4\eu^s}{(1+\eu^s)^2}\left[f_{\beta, k}-f_{\alpha, k} + \eu^s(f_{\alpha, k}+2f_{\beta, k}) \right].	
\end{align}
In Fig.~\ref{fig:coherentFidelity23(2)}, we plotted the second-order coefficient of the fidelity $F^{(2)}$ as a function of $u$ for $k=1,2,3$.  We observe from this figure that as the mode number $k$ increases, the fidelity decreases, which suggests that the optimal mode for encoding the quantum secret is $k=1$.\\  

In the limit $s\to \infty$, the fidelity up to third order is 
\begin{align}\label{sinfty}
F=1-4 \left( f_{\alpha, k} + 2f_{\beta, k} \right) h^2+O(h^3).
\end{align}
Hence, in the limit of infinite squeezing and in the absence of acceleration $(h=0)$, fidelity is one. However, for non-zero acceleration, fidelity is always smaller than one, even if a maximally entangled state is employed to encode the quantum secret.
\section{Conclusions and Discussions}\label{Discussion}
Here we study the effect of relativistic motion on $(2,3)$-threshold quantum Gaussian secret sharing. In our scheme, the dealer employs a single mode of a cavity to encode each quantum share. We begin by fully characterizing the BBB as a quantum Gaussian channel. We find that the canonical form of this channel is a thermal lossy channel. This form of the channel is useful for studying relativistic effects in quantum-information-processing tasks.\\

  We consider different possible collaboration scenarios between different subsets of players and analyze how each scenario can be written as a composition of quantum Gaussian channels. We find that the decoherence due to the relativistic motion of the quantum shares during distribution and also collaboration, reduces the fidelity of quantum secret sharing.\\ 
  
  Interestingly, we observe in the scenario wherein Players~1 and~2 are collaborating, depicted in Fig.~\ref{fig:collaboration12}, the fidelity is independent of the initial mean photon number in the encoded secret. Hence, in this case, the fidelity for a coherent state is the same as that of a vacuum state. Moreover, in the second scenario, Fig.~\ref{fig:collaboration23}, we find that when the quantum secret is a coherent state (or a vacuum state), the best encoding strategy is to encode the quantum secret in the ground mode of the cavity. We observe that the fidelity of the protocol is smaller than one, even in the limit of infinite squeezing, i.e., when maximal entanglement is used as a resource (See Eq.~\eqref{sinfty}).\\
  
  As a future line of research, we are interested in extending our results to the more general case of $(k,n)$-threshold quantum secret sharing. Furthermore, our hope is that the methods developed here can be employed to relax the conditions on spacetime replication of quantum states~\cite{Hayden1,Hayden2}, i.e., to consider the effect of non-uniform motion on this task.  
\acknowledgments
We thank Nicolai Friis, Masoud Habibi Davijani, and Abdullah Khalid for useful discussions and comments. The authors were supported by the Natural Sciences and Engineering Research Council of Canada (NSERC). BCS appreciates financial support from Alberta Innovates Technology Futures, China's 1000 Talent Plan and NSFC (Grant No. GG2340000241).

\bibliography{Relativisticquantumsecretsharing}

\end{document}